\documentclass[conference]{IEEEtran}
\IEEEoverridecommandlockouts

\usepackage{color}
\usepackage{bm}
\usepackage{graphicx}
\usepackage{amsmath}
\usepackage{amssymb}
\usepackage{multirow}
\usepackage{booktabs}
\usepackage{array}
\usepackage{amsthm}
\usepackage{lipsum}
\usepackage{enumerate}
\usepackage{stfloats}
\usepackage{subfigure}
\usepackage{cases}
\usepackage{halloweenmath}
\usepackage{cite}

\title{Wideband Modeling and Beamforming for Beyond Diagonal Reconfigurable Intelligent Surfaces}

\author{ \IEEEauthorblockN{Hongyu Li$^{\pumpkin}$, Matteo Nerini$^{\pumpkin}$, Shanpu Shen$^{\mathbat}$, and Bruno Clerckx$^{\pumpkin}$}

\IEEEauthorblockA{$^{\pumpkin}$ Department of Electrical and Electronic Engineering, Imperial College London, London SW7 2AZ, U.K. \\ E-mail: \texttt{\{c.li21,m.nerini20,b.clerckx\}@imperial.ac.uk} }

\IEEEauthorblockA{$^{\mathbat}$ Department of Electrical Engineering and Electronics, University of Liverpool, Liverpool L69 3GJ, U.K. \\ E-mail: \texttt{Shanpu.Shen@liverpool.ac.uk}  }
}

\begin{document}
\maketitle
\thispagestyle{empty}
\begin{abstract}
     This work studies the wideband modeling and beamforming design of beyond diagonal reconfigurable intelligent surface (BD-RIS), which generalizes and goes beyond conventional RIS with diagonal phase shift matrices to achieve enhanced channel gain. Specifically, we investigate the response of BD-RIS in wideband systems by going back to its hardware circuit realizations. We propose a novel wideband model which has simple expressions while capturing the response variations of BD-RIS for signals with different frequencies. With this wideband model, we propose a BD-RIS design algorithm for an orthogonal frequency division multiplexing system to maximize the average rate over all subcarriers. Finally, we provide simulation results to evaluate the performance of the proposed design and show the importance of wideband modeling for BD-RIS. 
\end{abstract}


\maketitle

\vspace{-0.1 cm}

\section{Introduction}

\vspace{-0.1 cm}

Beyond diagonal reconfigurable intelligent surface (BD-RIS) has been recently proposed to break through the limitation of conventional RIS with diagonal phase shift matrices \cite{wu2021intelligent} and provide enhanced channel gain and enlarged coverage \cite{li2023reconfigurable}. This is done by introducing inter-element connections, generating non-zero off-diagonal entries of the scattering matrix to provide more flexibility to BD-RIS for wave manipulation.
The concept of BD-RIS has been first proposed in \cite{shen2021}, where group- and fully-connected architectures have been introduced by interconnecting the RIS elements with additional tunable impedance components to enhance the channel strength compared to conventional RIS.
Other architectures, namely tree- and forest-connected architectures, have been proposed in \cite{nerini2023beyond} to reduce the circuit design complexity of BD-RIS while maintaining satisfactory performance.
In addition, BD-RISs supporting hybrid transmitting and reflecting and multi-sector modes have been proposed \cite{li2023reconfigurable}, which generalize and go beyond the intelligent omni-surface (IOS) \cite{zhang2022intelligent} to achieve full-space coverage while providing enhanced performance. 
While the aforementioned works \cite{li2023reconfigurable,shen2021,nerini2023beyond} model the BD-RIS aided communication system using the scattering parameter analysis, a universal framework bridging and characterizing the scattering, impedance, and admittance parameter-based RIS-aided communication models has been proposed in \cite{nerini2023universal}.

While the benefits of BD-RIS have been shown from multiple aspects \cite{li2023reconfigurable,shen2021,nerini2023beyond,nerini2023universal}, it is worth noticing that existing BD-RIS works are limited to narrowband communication systems. 
One recent work \cite{soleymani2024maximizing} has studied the beamforming design of BD-RIS in wideband communication systems, where the frequency independent BD-RIS model proposed for narrowband scenarios is directly adopted.
However, the scattering matrix of BD-RIS is frequency dependent and is determined by the circuit model of the reconfigurable impedance network. 
Therefore, how to accurately model the wideband BD-RIS and explore the benefits of BD-RIS in wideband communication systems remain an important open problem.
There are only limited works on the wideband modeling and beamforming design of conventional RIS \cite{li2021intelligent,jiang2021general,zhang2021joint,zheng2019intelligent}. The generalization to BD-RIS, however, is not straightforward due to the following twofold challenges. 
1) From the modeling perspective, in conventional RIS, each element is not connected to each other such that the phase and amplitude response for each element can be independently modeled based on specific circuit designs. Nevertheless, this does not hold for BD-RIS with interconnected elements, since the phase and amplitude of each entry of the scattering matrix depend also on those of other entries.  
2) From the beamforming perspective, the scattering matrix of BD-RIS has more intrincate constraints than conventional RIS coming from both the frequency dependency in the wideband model and the beyond diagonal property. 

To address the above two challenges, in this work, we explore the advantage of using admittance parameter analysis \cite{nerini2023universal} to simplify both the wideband modeling and beamforming design of BD-RIS. This leads to the following contributions.

\textit{First}, we propose a novel wideband model of BD-RIS based on lumped circuit designs. The proposed model has simple linear expressions while capturing accurately the frequency dependency of the BD-RIS admittance matrix. 
\textit{Second}, based on the proposed wideband BD-RIS model, we propose corresponding beamforming design algorithms to maximize the average rate for BD-RIS aided single-input single-output orthogonal frequency division multiplexing (SISO-OFDM) systems. 
\textit{Third}, we present simulation results to evaluate the effectiveness of the proposed beamforming design. The results also highlight the significance of investigating BD-RIS modeling in practical wideband communication systems.

\textit{Notations:}
$\mathbb{C}$ and $\mathbb{R}$ denote the set of complex and real numbers, respectively.
$(\cdot)^T$ and $(\cdot)^{-1}$ denote the transpose and inverse, respectively.
$\Re\{\cdot\}$ and $\Im\{\cdot\}$ denote the real and imaginary parts of a complex number, respectively.
$\mathsf{blkdiag}(\cdot)$ denotes a block-diagonal matrix.
$\mathsf{vec}(\cdot)$ and $\overline{\mathsf{vec}}(\cdot)$ denote the vectorization and its reverse operation, respectively.
$[\mathbf{a}]_{i:j}$ extracts the $i$-th through the $j$-th entries of $\mathbf{a}$.

\section{Wideband Modeling of BD-RIS}

In this section, we first review the BD-RIS modeling using admittance parameter analysis, and then establish the wideband BD-RIS model by characterizing its frequency dependency based on the circuit model.

\subsection{Admittance Matrix of BD-RIS}

\begin{figure}
    \centering
    \includegraphics[width=0.47\textwidth]{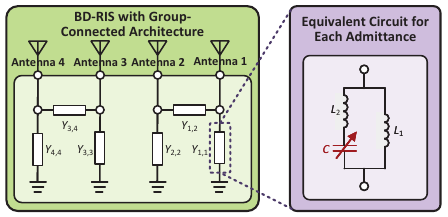}
    \caption{Example of an 4-element BD-RIS with group-connected reconfigurable admittance network and the equivalent circuit for each admittance.}\label{fig:bdris_cir}
\end{figure}

Consider an $M$-element BD-RIS modeled as $M$ antennas connected to an $M$-port reconfigurable admittance network \cite{shen2021}. Specifically, for a general group-connected reconfigurable admittance network with $G$ uniform groups, the $\bar{M} = M/G$ ports in each group are connected to each other by tunable admittance components, while ports from different groups are not connected to each other \cite{shen2021,nerini2023beyond}. 
To facilitate understanding, an example of a 4-element BD-RIS with group-connected architectures is illustrated in Fig. \ref{fig:bdris_cir}.  
According to the circuit topology of the group-connected reconfigurable admittance network, the corresponding admittance matrix $\mathbf{Y}\in\mathbb{C}^{M\times M}$ has a block-diagonal structure defined as 
\begin{equation}
    \mathbf{Y} = \mathsf{blkdiag}(\mathbf{Y}_1,\ldots,\mathbf{Y}_G),
    \label{eq:Y_blkdiag}
\end{equation}
where $\mathbf{Y}_g\in\mathbb{C}^{\bar{M}\times\bar{M}}$, $\forall g\in\mathcal{G} = \{1,\ldots,G\}$ is symmetric and purely imaginary for reciprocal and lossless reconfigurable admittance networks, that is 
\begin{equation}
    \mathbf{Y}_g = \mathbf{Y}_g^T, ~\Re\{\mathbf{Y}_g\} = \mathbf{0}_{\bar{M}}, \forall g.
    \label{eq:Y_symmetric}
\end{equation}
In group $g$ of the group-connected reconfigurable admittance network, each port $m_{g} = (g-1)\bar{M}+m$ is connected to port $m'_g = (g-1)\bar{M}+m'$, with a tunable admittance\footnote{$Y_{m_g,m_g}$ refers to the admittance which connects port $m_g$ to ground.} $Y_{m_g,m'_g}$, $\forall m,m'\in\bar{\mathcal{M}}=\{1,\ldots,\bar{M}\}$, which satisfies $Y_{m_g,m'_g} = Y_{m'_g,m_g}$ for $m'>m$. 
Accordingly, $\mathbf{Y}_g, \forall g\in\mathcal{G}$ writes as \cite{nerini2023beyond}
\begin{equation}
    [\mathbf{Y}_g]_{m,m'} = \begin{cases}
        -Y_{m_g,m'_g}, &m'\ne m \\
        \sum_{k\in\bar{\mathcal{M}}}Y_{m_g,k_g}, & m' = m
    \end{cases}, \forall m,m',g.
    \label{eq:Y_calculate}
\end{equation}

In equations (\ref{eq:Y_blkdiag})-(\ref{eq:Y_calculate}), we have general expressions of the admittance matrix for BD-RIS without specifying its frequency dependency, which will be identified in the following subsection based on the circuit model of each tunable admittance.

\subsection{Wideband Modeling of Tunable Admittance}

\begin{figure}
    \centering
    \subfigure[Susceptance versus frequency]{
    \includegraphics[width=0.23\textwidth]{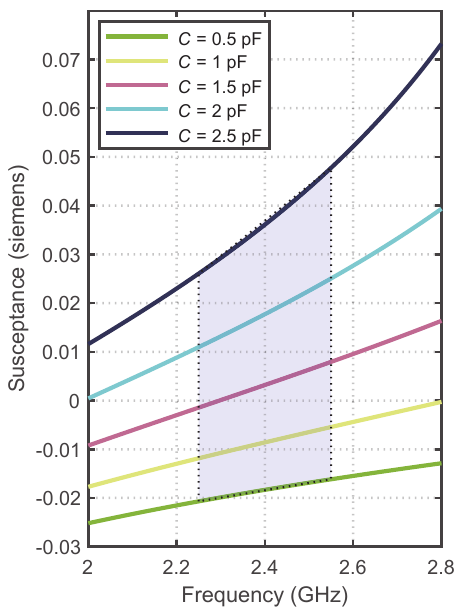}}
    \subfigure[Susceptance with fitted model]{
        \includegraphics[width=0.23\textwidth]{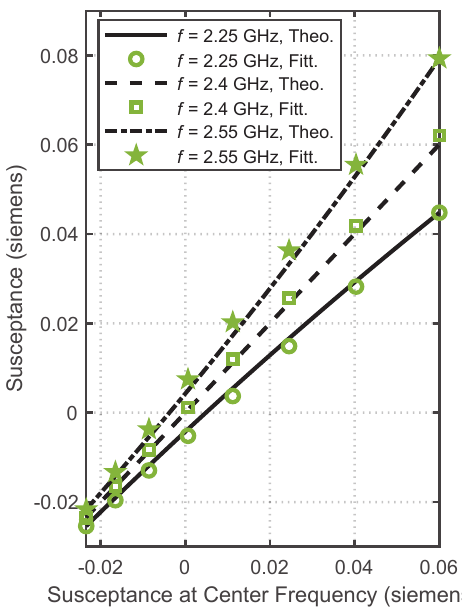}}
    \caption{The susceptance as a function of (a) frequency and (b) the value of susceptance at the center frequency $f_\mathrm{c} = 2.4$ GHz with $L_1 = 2.5$ nH, $L_2 = 0.7$ nH, and $C\in[0.2,3]$ pF for a practical varactor diode. The parameters $\alpha_j$, $\beta_j$, $\forall j\in\{1,2\}$ in the linear model (\ref{eq:fitted_model}) are set as $\alpha_1 = 1.2161\times10^{-9}$, $\alpha_2 = -1.9076$, $\beta_1 = 4.0925\times10^{-11}$, $\beta_2 = -0.098$.}
    \label{fig:susceptance_f}
\end{figure}

In this work, we model each tunable admittance component as a lumped circuit consisting of two inductors $L_1$ and $L_2$, and one tunable capacitor $C$ \cite{koziel2013surrogate}, as illustrated in Fig. \ref{fig:bdris_cir}\footnote{Here we remove the equivalent resistance in the circuit to make the admittance network lossless. Detailed studies on the impact of lossy admittance networks can be found in \cite{nerini2024localized}.}. 
The admittance of the circuit is thus a function of $C$ and the frequency for the incident signals, that is\footnote{In this model, we set fixed and same values of inductors $L_1$ and $L_2$ for each admittance component, while making the value of the capacitor $C$ tunable and different, i.e., $C_{m_g,m'_g}$, for different admittance components.}
\begin{equation}
    Y_{m_g,m'_g}(C,f) = \frac{1}{\jmath2\pi fL_1} + \frac{1}{\jmath 2\pi f L_2 + \frac{1}{\jmath 2\pi fC}}, \forall m,m',g,
    \label{eq:admittance_C_f}
\end{equation}
yielding the corresponding susceptance as 
$B_{m_g,m'_g}(C,f) = \Im\{Y_{m_g,m'_g}(C,f)\}$, $\forall m,m'\in\bar{\mathcal{M}}$, $\forall g\in\mathcal{G}$.
From (\ref{eq:admittance_C_f}) we can observe that the tunable admittance values at different frequencies are coupled to each other, that is for a certain value of capacitor $C$, the admittance will have different values at different frequencies. 
More importantly, this is an intrinsic phenomenon coming from the hardware implementation of BD-RIS elements, which cannot be simply ignored in BD-RIS aided wideband communication systems. 
This phenomenon is also numerically illustrated in Fig. \ref{fig:susceptance_f}(a), where we consider a wideband system with center frequency $f_\mathrm{c} = 2.4$ GHz and adopt a practical varactor diode according to the data sheet for the diode SMV1231-079.  
From Fig. \ref{fig:susceptance_f}(a) we observe that the value of the susceptance varies faster with frequency for increasing $C$. 
In addition, the susceptance can be regarded as a linear function with respect to frequency for different values of $C$ within some practical bandwidth for wideband communication systems, such as the range from 2.25 GHz to 2.55 GHz as highlighted in Fig. \ref{fig:susceptance_f}(a). 
This motivates us to model the frequency dependency of each admittance component for BD-RIS in wideband communication systems as a simplified linear function, which is given by:
\begin{subequations}\label{eq:fitted_model}
    \begin{align} 
        \nonumber
        &B_{m_g,m'_g}(B_{\mathrm{c},m_g,m'_g},f) \\
        &~~~~~~~= F_1(f)B_{\mathrm{c},m_g,m'_g} + F_2(f), \forall m,m',g,\\
        &F_1(f) = \alpha_1 f + \beta_1, ~~
        F_2(f) = \alpha_2 f + \beta_2,
    \end{align}
\end{subequations} 
where $B_{\mathrm{c},m_g,m'_g}$ denotes the value of the susceptance at the center frequency $f_\mathrm{c}$, parameters $\alpha_j$, $\beta_j$, $\forall j\in\{1,2\}$ are determined by $L_1$, $L_2$, and $f_\mathrm{c}$\footnote{In this work, we obtain the values of $\alpha_j$, $\beta_j$, $\forall j\in\{1,2\}$ by linearly fitting the trend of (\ref{eq:admittance_C_f}) as a function of $B_{\mathrm{c},m_g,m'_g}$ (calculated by setting $f = f_\mathrm{c}$ in (\ref{eq:admittance_C_f})), where both the slope and intercept are regarded as linear functions of frequency such that each of them can be characterized based on arbitrary two points on the lines.}. 
For the highlighted curves in Fig. \ref{fig:susceptance_f}(a), the theoretical and fitted results using the linear model (\ref{eq:fitted_model}) are shown in Fig.  \ref{fig:susceptance_f}(b). The results guarantee a $0.27\%$ normalized mean square error between the fitted values and theoretical ones, which demonstrate the accuracy of the proposed linear model.
In the following section, we will adopt this model in a BD-RIS aided wideband communication system and design the beamforming for BD-RIS.

\section{Beamforming Design for \\BD-RIS Aided SISO-OFDM}

In this section, we apply the proposed wideband BD-RIS model in a SISO-OFDM system, formulating the rate maximization problem, and developing corresponding algorithms to optimize BD-RIS.

\subsection{Problem Formulation}

We consider a BD-RIS aided wideband SISO-OFDM system with $N$ subcarriers. Denoting $s_n\in\mathbb{C}$ the transmit symbol associated with subcarrier $n$, $\mathbb{E}\{|s_n|^2\} = 1$, $\forall n\in\mathcal{N} = \{1,\ldots,N\}$, the received signal for subcarrier $n$ is given by 
$y_n = \sqrt{p_n}h_ns_n + z_n, \forall n\in\mathcal{N}$,
where $p_n\ge0$ denotes the transmit power allocated to symbol $s_n$, $\sum_{n\in\mathcal{N}}p_n\le P$ with total transmit power $P$, $z_n\in\mathbb{C}$ denotes the noise at the receiver with $z_n\thicksim\mathcal{CN}(0,\sigma^2)$, and $h_n\in\mathbb{C}$ denotes the frequency-domain BD-RIS aided wireless channel between the transmitter and the receiver at subcarrier $n$. 
According to \cite{nerini2023universal}, by modeling the transmitter, BD-RIS, and receiver as a multi-port network, the wireless channel can be characterized by its admittance matrix.
Specifically, the BD-RIS aided channel $h_n$ can be formulated as \cite{nerini2023universal}
\begin{equation}
    h_n = \frac{1}{2Y_0}(-y_{RT,n} + \mathbf{y}_{RI,n}(\bar{\mathbf{Y}}_n + Y_0\mathbf{I}_M)^{-1}\mathbf{y}_{IT,n}), \forall n,
    \label{eq:channel}
\end{equation}
where $Y_0$ refers to the characteristic admittance, e.g., $\frac{1}{50}$ S, $y_{RT,n}\in\mathbb{C}$, $\mathbf{y}_{RI,n}\in\mathbb{C}^{1\times M}$, and $\mathbf{y}_{IT,n}\in\mathbb{C}^{M\times 1}$, respectively, refer to the channels from transmitter to receiver, from BD-RIS to receiver, and from transmitter to BD-RIS based on the admittance parameters \cite{nerini2023universal}. According to the discussions in \cite{nerini2023universal}, these channels are a characterization of the wireless channels equivalent to the widely used models based on the scattering parameters \cite{wu2021intelligent}. The detailed mapping between the channels based on the scattering parameters and those based on the admittance parameters can be found in \cite{nerini2023universal}, and will be briefly summarized in Section IV.
In addition, $\bar{\mathbf{Y}}_n\in\mathbb{C}^{M\times M}$ refers to the admittance matrix of BD-RIS at subcarrier $n$, which is constructed based on equations (\ref{eq:Y_blkdiag})-(\ref{eq:Y_calculate}) and the linear model (\ref{eq:fitted_model}) of each tunable admittance component. Specifically, given the carrier frequency $f_n$ for subcarrier $n$, $\bar{\mathbf{Y}}_n$ is
\begin{subequations}\label{eq:Y_wideband_construction}
    \begin{align}
        \label{eq:blkdiag}
        &\bar{\mathbf{Y}}_n = \mathsf{blkdiag}(\mathbf{Y}_{1,n},\ldots,\mathbf{Y}_{G,n}), \forall n, \\
        \label{eq:symmetric}
        &\mathbf{Y}_{g,n} = \jmath\mathbf{B}_{g,n}, ~\mathbf{B}_{g,n} = \mathbf{B}_{g,n}^T, \forall g,n,\\
        \label{eq:B_cal}
        &[\mathbf{B}_{g,n}]_{m,m'} = \begin{cases}
            -B_{m_g,m'_g,n}, &m'\ne m \\
            \sum_{k\in\bar{\mathcal{M}}}B_{m_g,k_g,n}, & m' = m
        \end{cases},\\
        \label{eq:wideband_constraint}
        &B_{m_g,m'_g,n} = F_1(f_n)B_{\mathrm{c},m_g,m'_g} + F_2(f_n), \forall m,m',g,n,\\
        \label{eq:range}
        &B_{\mathrm{c},m_g,m'_g} \in[B_\mathrm{min},B_\mathrm{max}], \forall m,m',g,
    \end{align}
\end{subequations}
where $B_\mathrm{min}$ and $B_\mathrm{max}$ are determined by the practical range of capacitor values. 
From (\ref{eq:Y_wideband_construction}) we observe that $\bar{\mathbf{Y}}_n$ for all carrier frequencies are dependent to each other and jointly tuned by $B_{\mathrm{c},m_g,m'_g}$. Therefore, we aim to jointly design the power allocation at the transmitter and the susceptances $B_{\mathrm{c},m_g,m'_g}$ at the center frequency to maximize the average rate for the SISO-OFDM system. This yields the following problem:
\begin{equation}\label{eq:prob0}
    \begin{aligned}
    &\{\{B_{\mathrm{c},m_g,m'_g}^\star\}_{\forall m,m',g}, \{p_n^\star\}_{\forall n}\}\\
    &~~~= \arg~\max_{\text{(\ref{eq:Y_wideband_construction})},\sum_n p_n\le P}\frac{1}{N}\sum_{n\in\mathcal{N}}\log_2\Big(1+\frac{p_n|h_n|^2}{\sigma^2}\Big).
    \end{aligned}
\end{equation} 

\subsection{Optimization}

Problem (\ref{eq:prob0}) is a challenging joint approximation mainly due to the coupling constraints of BD-RIS in wideband systems. 
To simplify the design, we decouple problem (\ref{eq:prob0}) into two sub-problems and solve each of them individually. 

\subsubsection{BD-RIS Design}
Since the constraint (\ref{eq:wideband_constraint}) of BD-RIS coming from the wideband modeling couples the admittances for different subcarriers, which are embedded in the $\log(\cdot)$ function, it is difficult to directly design the BD-RIS with the original objective function in problem (\ref{eq:prob0}). This motivates us to first simplify the objective function in (\ref{eq:prob0}) by using the Jensen's inequality, which yields
$\frac{1}{N}\sum_{n\in\mathcal{N}}\log_2(1+\frac{p_n|h_n|^2}{\sigma^2}) \le \log_2(1+\frac{1}{N}\sum_{n\in\mathcal{N}}\frac{p_n|h_n|^2}{\sigma^2})$.
In addition, we remove the impact of power allocation when designing BD-RIS by setting $p_1 = \cdots = p_N$, while the power allocation will be designed after the admittance matrices of BD-RIS have been determined.
This leads to the following sum channel gain maximization problem:
\begin{equation}
    \begin{aligned}
        &\{B_{\mathrm{c},m_g,m'_g}^\star\}_{\forall m,m',g}
        = \arg\max_{\text{(\ref{eq:symmetric})-(\ref{eq:range})}} \sum_{n\in\mathcal{N}}\Big|-y_{RT,n}\\
        &~~~~~~~~~+\sum_{g\in\mathcal{G}}\mathbf{y}_{RI,g,n}(\jmath\mathbf{B}_{g,n}+Y_0\mathbf{I}_{\bar{M}})^{-1}\mathbf{y}_{IT,g,n}\Big|^2,
    \end{aligned}\label{eq:sub_prob1}
\end{equation}
where $\mathbf{y}_{RI,g,n} = [\mathbf{y}_{RI,n}]_{(g-1)\bar{M}+1:g\bar{M}}$, and $\mathbf{y}_{IT,g,n} = [\mathbf{y}_{IT,n}]_{(g-1)\bar{M}+1:g\bar{M}}$, $\forall g\in\mathcal{G}$. We will transform (\ref{eq:sub_prob1}) into an unconstrained optimization based on the following steps.

\textit{Step 1: Eliminating Constraints (\ref{eq:symmetric}) And (\ref{eq:B_cal})}.
We start by re-describing (\ref{eq:B_cal}) as a function which maps the matrix $\bar{\mathbf{B}}_{g,n}\in\mathbb{C}^{\bar{M}\times\bar{M}}$ with entries $[\bar{\mathbf{B}}_{g,n}]_{m,m'} = B_{m_g,m'_g,n}$ to the matrix $\mathbf{B}_{g,n}$, i.e., $\mathbf{B}_{g,n} = F_3(\bar{\mathbf{B}}_{g,n})$, where $F_3(\cdot)$ is given such that $[\mathbf{B}_{g,n}]_{m,m'} = -[\bar{\mathbf{B}}_{g,n}]_{m,m'}$ for $m'\ne m$ and $[\mathbf{B}_{g,n}]_{m,m} = \sum_{k\in\bar{\mathcal{M}}}[\bar{\mathbf{B}}_{g,n}]_{m,k}$.
In addition, the symmetric constraint of each $\mathbf{B}_{g,n}$ in (\ref{eq:symmetric}) and the linear mapping (\ref{eq:B_cal}) imply that each $\bar{\mathbf{B}}_{g,n}$ should be symmetric, i.e., $\bar{\mathbf{B}}_{g,n} = \bar{\mathbf{B}}_{g,n}^T$. 
In other words, each $\bar{\mathbf{B}}_{g,n}$ is essentially determined by its $\bar{M}$ diagonal and $\frac{\bar{M}(\bar{M}-1)}{2}$ lower-triangular (or upper-triangular) entries. 
This motivates us to rewrite $\bar{\mathbf{B}}_{g,n}$ as 
$\bar{\mathbf{B}}_{g,n} = \overline{\mathsf{vec}}(\mathbf{P}\bar{\mathbf{b}}_{g,n})$, where
$\bar{\mathbf{b}}_{g,n}\in\mathbb{R}^{\frac{\bar{M}(\bar{M}+1)}{2}\times 1}$ contains the diagonal and lower-triangular entries of $\bar{\mathbf{B}}_{g,n}$.
$\mathbf{P}\in\{0,1\}^{\bar{M}^2\times\frac{\bar{M}(\bar{M}+1)}{2}}$ is the binary matrix mapping $\bar{\mathbf{b}}_{g,n}$ into $\mathsf{vec}(\bar{\mathbf{B}}_{g,n})$, $\forall g\in\mathcal{G}$ defined as 
\begin{equation}
    \label{eq:p}
    [\mathbf{P}]_{\bar{M}(i-1)+i',l} = \begin{cases}
        1, & l = \frac{i(i-1)}{2}+i' ~\text{and}~ 1 \le i' \le i,\\
        1, & l = \frac{i'(i'-1)}{2}+i ~\text{and}~ i<i'\le \bar{M},\\
        0, & \text{otherwise},
    \end{cases}
\end{equation}
where $\forall i, i' \in\bar{\mathcal{M}}$.
According to (\ref{eq:wideband_constraint}), these vectors $\bar{\mathbf{b}}_{g,n}$ for all subcarriers are jointly controlled by $\bar{\mathbf{b}}_{\mathrm{c},g}\in\mathbb{R}^{\frac{\bar{M}(\bar{M}+1)}{2}\times 1}$ constructed with the entries $B_{\mathrm{c},m_g,m'_g}$ for $m'\ge m$.
Therefore, we can reformulate problem (\ref{eq:sub_prob1}) as the following form:
\begin{subequations}\label{eq:sub_prob2}
    \begin{align}
    \nonumber
    \max_{\{\bar{\mathbf{b}}_{\mathrm{c},g\}_{\forall g}}} &\sum_{n\in\mathcal{N}}\Big|-y_{RT,n}+\sum_{g\in\mathcal{G}}\mathbf{y}_{RI,g,n}\\
    &~~\times(\jmath F_3(\overline{\mathsf{vec}}(\mathbf{P}\bar{\mathbf{b}}_{g,n}))+Y_0\mathbf{I}_{\bar{M}})^{-1}\mathbf{y}_{IT,g,n}\Big|^2\\
    \label{eq:constraint_wideband}
    \text{s.t.} ~~&\bar{\mathbf{b}}_{g,n} = F_1(f_n)\bar{\mathbf{b}}_{\mathrm{c},g} + F_2(f_n), \forall n,g,\\
    \label{eq:constraint_range}
    &[\bar{\mathbf{b}}_{\mathrm{c},g}]_l \in [B_\mathrm{min},B_\mathrm{max}], \forall g, l\in\mathcal{L},
    \end{align}
\end{subequations}
where $\mathcal{L} = \{1,\ldots,\frac{\bar{M}(\bar{M}+1)}{2}\}$.

\textit{Step 2: Eliminating Constraint (\ref{eq:constraint_range})}.
The linear constraint (\ref{eq:constraint_range}) of each entry of $\bar{\mathbf{b}}_{\mathrm{c},g}$ can be further removed by one-to-one mapping variables with arbitrary real values into the range $[B_\mathrm{min},B_\mathrm{max}]$. This can be done by introducing a tractable mapping function $[\bar{\mathbf{b}}_{\mathrm{c},g}]_l = F_4([\mathbf{x}_g]_l)$ with $\mathbf{x}_g\in\mathbb{R}^{\frac{\bar{M}(\bar{M}+1)}{2}\times 1}$, $\forall l\in\mathcal{L}$, such that 
\begin{enumerate}[$\circ$]
    \item $F_4([\mathbf{x}_g]_l)$ is continuous, differentiable, and monotonously increasing in the full range $[\mathbf{x}_g]_l\in(-\infty,+\infty)$;
    \item $\lim_{[\mathbf{x}_g]_l\rightarrow +\infty}F_4([\mathbf{x}_g]_l) = B_\mathrm{max}$;
    \item $\lim_{[\mathbf{x}_g]_l\rightarrow -\infty}F_4([\mathbf{x}_g]_l) = B_\mathrm{min}$.
\end{enumerate}
An example of such a mapping function is given by:
\begin{equation}
    [\bar{\mathbf{b}}_{\mathrm{c},g}]_l = F_4([\mathbf{x}_g]_l) = \frac{[\mathbf{x}_g]_l}{\sqrt{[\mathbf{x}_g]_l^2B_-^{-2} + 1}} + B_+, \forall g, l,
    \label{eq:mapping}
\end{equation}
where $B_- = \frac{B_\mathrm{max}-B_\mathrm{min}}{2}$ and $B_+ = \frac{B_\mathrm{max}+B_\mathrm{min}}{2}$. 
By this mapping function, we can transform problem (\ref{eq:sub_prob2}) into the following unconstrained optimization: 
\begin{equation}\label{eq:sub_prob3}
    \begin{aligned}
        \{\mathbf{x}_g^\star\}_{\forall g} = &\arg\max_{\text{(\ref{eq:constraint_wideband}), (\ref{eq:mapping})}} \sum_{n\in\mathcal{N}}\Big|-y_{RT,n}+\sum_{g\in\mathcal{G}}\mathbf{y}_{RI,g,n}\\
        &~~\times(\jmath F_3(\overline{\mathsf{vec}}(\mathbf{P}\bar{\mathbf{b}}_{g,n}))+Y_0\mathbf{I}_{\bar{M}})^{-1}\mathbf{y}_{IT,g,n}\Big|^2,
    \end{aligned}
\end{equation}
which can be directly solved by the quasi-Newton method. 

With the solutions $\mathbf{x}_{g}^\star$ to problem (\ref{eq:sub_prob3}), we can obtain $\bar{\mathbf{b}}_{\mathrm{c},g}^\star$ by (\ref{eq:mapping}), $\bar{\mathbf{b}}_{g,n}^\star$ by (\ref{eq:constraint_wideband}), and $\mathbf{B}_{g,n}^\star = F_3(\overline{\mathsf{vec}}(\mathbf{P}\bar{\mathbf{b}}_{g,n}^\star))$, $\forall g\in\mathcal{G}$, $\forall n\in\mathcal{N}$, such that the effective channel $h_n^\star$ can be obtained by (\ref{eq:channel}) with $\bar{\mathbf{Y}}_n^\star = \jmath\mathsf{blkdiag}(\mathbf{B}_{1,n}^\star,\ldots,\mathbf{B}_{G,n}^\star)$.

\subsubsection{Power Allocation}
When the BD-RIS aided wideband channels are determined, problem (\ref{eq:prob0}) boils down to the conventional power allocation problem,
which can be solved by the well-known water-filling method.

\section{Performance Evaluation}

In this section, we evaluate the performance of the BD-RIS aided SISO-OFDM system to show the importance of the wideband modeling for BD-RIS.

\subsection{Simulation Setup}
In the considered SISO-OFDM system, we assume the center frequency as $f_\mathrm{c} = 2.4$ GHz, the bandwidth as 300 MHz, and the number of subcarriers as $N=64$. 
The OFDM channels are modeled with the maximum delay spread of 16 taps using circularly symmetric complex Gaussian random variables.  
The length of the cyclic prefix is set to be 16. 
The distance-dependent channel pathloss is modeled as $\zeta_o = \zeta_0d_o^{-\varepsilon_o}$, $\forall o\in\{RT,RI,IT\}$, where $\zeta_0 = -30$ dB refers to the signal attenuation at reference distance 1 m, $d_o$ denotes the distance between transmitter/BD-RIS/receiver, and $\varepsilon_o$ denotes the pathloss exponent. In the following simulations, we set $d_{RT} = 33$ m, $d_{RI} = 5$ m, $d_{IT} = 30$ m, $\varepsilon_{RT} = 3.8$, $\varepsilon_{RI} = 2.2$, and $\varepsilon_{IT} = 2.5$.
The noise power is set as $\sigma^2 = -80$ dBm. 
With the frequency-domain channels modeled from above, i.e., $h_{RT,n}\in\mathbb{C}$, $\mathbf{h}_{RI,n}\in\mathbb{C}^{1\times M}$, and $\mathbf{h}_{IT,n}\in\mathbb{C}^{M\times 1}$, $\forall n\in\mathcal{N}$, we can obtain the corresponding channels based on the admittance parameters by \cite{nerini2023universal}
\begin{equation}
    \begin{aligned}
        &\mathbf{y}_{RI,n} = -2Y_0\mathbf{h}_{RI,n}, ~~ \mathbf{y}_{IT,n} = -2Y_0\mathbf{h}_{IT,n},\\
        &y_{RT,n} = -2Y_0(h_{RT,n} - \mathbf{h}_{RI,n}\mathbf{h}_{IT,n}), \forall n.
    \end{aligned}
\end{equation}  

\subsection{Benchmark Scheme}
To show how the wideband modeling of BD-RIS impacts the system performance, we include the case where the BD-RIS admittances are designed such that they remain the same for all subcarriers. 
That is, we ignore the constraint (\ref{eq:wideband_constraint}), which yields $\bar{\mathbf{Y}}_1 = \ldots = \bar{\mathbf{Y}}_N = \mathbf{Y}$ as in \cite{soleymani2024maximizing} where the narrowband BD-RIS model is adopted in wideband scenarios. We have 
\begin{equation}
    \bar{h}_n = \frac{1}{2Y_0}(-y_{RT,n} + \mathbf{y}_{RI,n}(\mathbf{Y} + Y_0\mathbf{I}_M)^{-1}\mathbf{y}_{IT,n}), \forall n.
\end{equation}  
Then the proposed algorithm can also be adopted to solve the following problem:
\begin{equation}
    \begin{aligned}
    &\{\{Y_{m_g,m'_g}^\sharp\}_{\forall m,m',g}, \{p_n^\sharp\}_{\forall n}\}\\
    &~~ = \mathop{\mathrm{arg~~~max}}\limits_{\substack{\text{(\ref{eq:Y_blkdiag})-(\ref{eq:Y_calculate})}, \sum_{n\in\mathcal{N}} p_n\le P\\ \Im\{Y_{m_g,m'_g}\}\in[B_\mathrm{min},B_\mathrm{max}]}} ~\frac{1}{N}\sum_{n\in\mathcal{N}}\log_2\Big(1+\frac{p_n|\bar{h}_n|^2}{\sigma^2}\Big).
    \end{aligned}
\end{equation} 
The obtained solutions $\{Y_{m_g,m'_g}^\sharp\}_{\forall m,m',g}$ and $\{p_n^\sharp\}_{\forall n}$ are further plugged into the original objective function including the wideband effect (\ref{eq:wideband_constraint}) for evaluation.

\subsection{System Performance}

\begin{figure}
    \centering
    \includegraphics[width=0.44\textwidth]{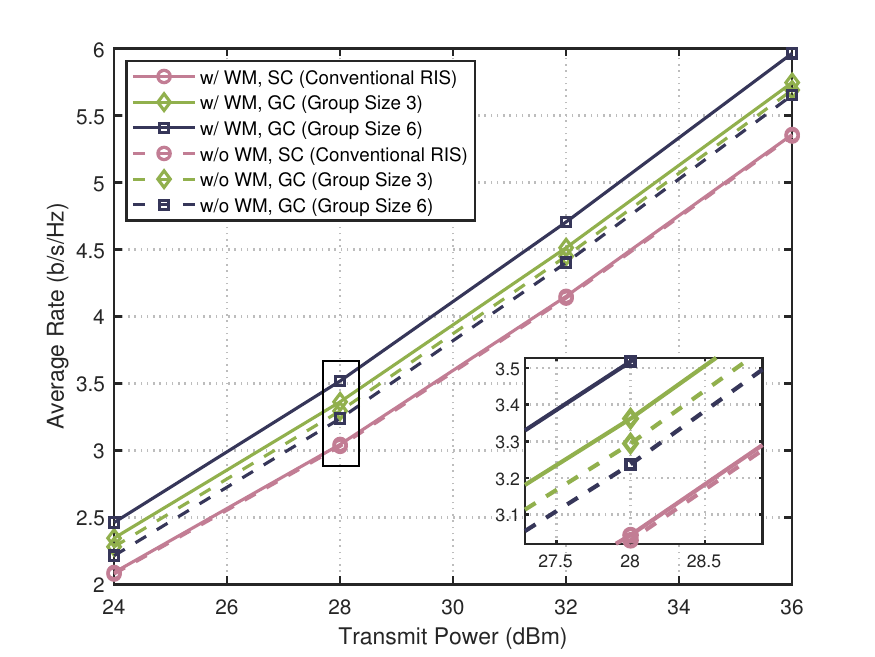}
    \caption{Average rate versus transmit power $P$ with BD-RIS having different architectures ($M = 36$, $\bar{M}\in\{1,3,6\}$). The legend ``WM'' is short for wideband modeling; ``SC'' is short for single-connected ($\bar{M}=1$); ``GC'' is short for group-connected ($\bar{M}\in\{3,6\}$).}
    \label{fig:AR_P}\vspace{-0.2 cm}
\end{figure}

\begin{figure}
    \centering
    \includegraphics[width=0.44\textwidth]{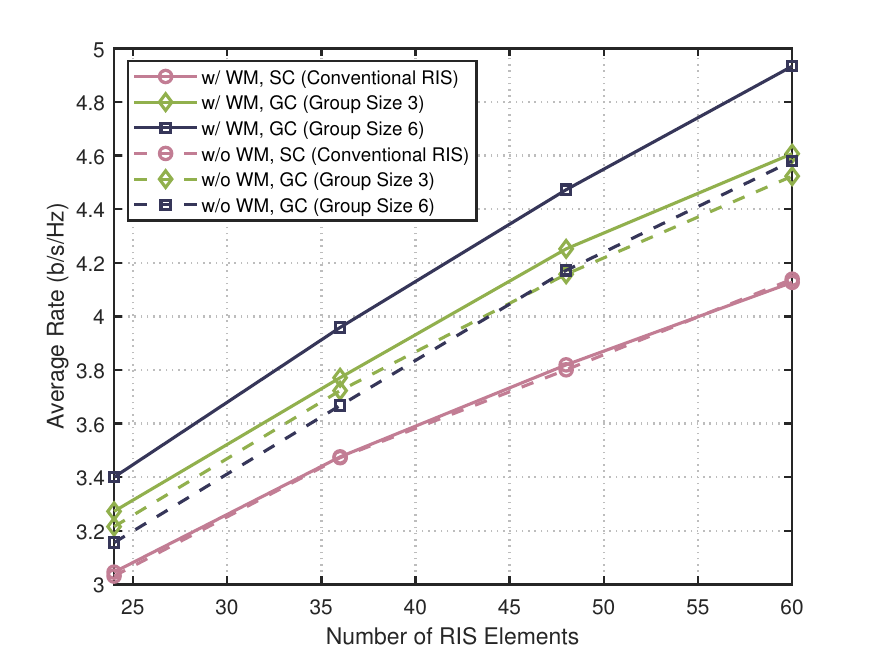}
    \caption{Average rate versus the number of RIS elements $M$ with BD-RIS having different architectures ($P = 30$ dBm, $\bar{M}\in\{1,3,6\}$).}\vspace{-0.2 cm}
    \label{fig:AR_M}
\end{figure}

We evaluate the performance of the proposed wideband modeling and algorithm design by plotting 
the average rate of the BD-RIS aided SISO-OFDM system versus, respectively, the transmit power $P$ in Fig. \ref{fig:AR_P} and the number of RIS elements $M$ in Fig. \ref{fig:AR_M}. 
From Figs. \ref{fig:AR_P} and \ref{fig:AR_M}, we have the following observations. 
\textit{First}, when taking into account the wideband modeling, BD-RIS with larger group size of the group-connected architecture achieves better performance within all considered ranges of transmit power and number of RIS elements. 
This is attributed to the increasing wave manipulation capabilities offered by more complex circuit topologies of BD-RIS. 
\textit{Second}, when ignoring the wideband effect for designing BD-RIS, BD-RIS architectures with larger group size cannot always provide better performance. 
For example, BD-RIS with group size $\bar{M}=3$ outperforms that with $\bar{M} = 6$ when $M = 36$, while the latter outperforms the former for $M=60$.
This is because the variation for BD-RIS matrices between different subcarriers becomes more significant with increasing number of admittances, such that directly designing BD-RIS ignoring this variation will lead to increasing performance loss. 
\textit{Third}, the performance gap between the above two schemes becomes significant on BD-RIS architectures with a larger $\bar{M}$, demonstrating the importance of accurately modeling BD-RIS in wideband systems.

\section{Conclusion}
\label{sc:Conclusion}

In this paper, we model BD-RIS in wideband communication systems based on the circuit realizations of the reconfigurable admittance network. 
The proposed model has simple expressions while it can accurately capture the frequency dependency of the admittance matrix for BD-RIS. 
With the proposed wideband model, we design the beamforming to maximize the average rate for a BD-RIS aided SISO-OFDM system. 
We finally present simulation results which verified the effectiveness of the proposed beamforming design and demonstrate the importance of modeling the response of BD-RIS in wideband wireless systems.

\bibliographystyle{IEEEbib}
\bibliography{refs}

\begin{thebibliography}{10}

\bibitem{wu2021intelligent}
Qingqing Wu, Shuowen Zhang, Beixiong Zheng, Changsheng You, and Rui Zhang,
\newblock ``Intelligent reflecting surface-aided wireless communications: A
  tutorial,''
\newblock {\em IEEE Trans. Commun.}, vol. 69, no. 5, pp. 3313--3351, 2021.

\bibitem{li2023reconfigurable}
Hongyu Li, Shanpu Shen, Matteo Nerini, and Bruno Clerckx,
\newblock ``Reconfigurable intelligent surfaces 2.0: Beyond diagonal phase
  shift matrices,''
\newblock {\em IEEE Commun. Mag.}, vol. 62, no. 3, pp. 102--108, 2024.

\bibitem{shen2021}
Shanpu Shen, Bruno Clerckx, and Ross Murch,
\newblock ``Modeling and architecture design of reconfigurable intelligent
  surfaces using scattering parameter network analysis,''
\newblock {\em IEEE Trans. Wireless Commun.}, vol. 21, no. 2, pp. 1229--1243,
  2021.

\bibitem{nerini2023beyond}
Matteo Nerini, Shanpu Shen, Hongyu Li, and Bruno Clerckx,
\newblock ``Beyond diagonal reconfigurable intelligent surfaces utilizing graph
  theory: Modeling, architecture design, and optimization,''
\newblock {\em IEEE Trans. Wireless Commun.}, 2024.

\bibitem{zhang2022intelligent}
Hongliang Zhang, Shuhao Zeng, Boya Di, Yunhua Tan, Marco Di~Renzo, M{\'e}rouane
  Debbah, Zhu Han, H~Vincent Poor, and Lingyang Song,
\newblock ``Intelligent omni-surfaces for full-dimensional wireless
  communications: Principles, technology, and implementation,''
\newblock {\em IEEE Commun. Mag.}, vol. 60, no. 2, pp. 39--45, 2022.

\bibitem{nerini2023universal}
Matteo Nerini, Shanpu Shen, Hongyu Li, and Bruno Clerckx,
\newblock ``A universal framework for multiport network analysis of
  reconfigurable intelligent surfaces,''
\newblock {\em arXiv:2311.10561}, 2023.

\bibitem{soleymani2024maximizing}
Mohammad Soleymani, Ignacio Santamaria, Aydin Sezgin, and Eduard Jorswieck,
\newblock ``Maximizing spectral and energy efficiency in multi-user {MIMO OFDM}
  systems with {RIS} and hardware impairment,''
\newblock {\em arXiv preprint arXiv:2401.11921}, 2024.

\bibitem{li2021intelligent}
Hongyu Li, Wenhao Cai, Yang Liu, Ming Li, Qian Liu, and Qingqing Wu,
\newblock ``Intelligent reflecting surface enhanced wideband {MIMO-OFDM}
  communications: From practical model to reflection optimization,''
\newblock {\em IEEE Trans. Commun.}, vol. 69, no. 7, pp. 4807--4820, 2021.

\bibitem{jiang2021general}
Hao Jiang, Chengyao Ruan, Zaichen Zhang, Jian Dang, Liang Wu, Mithun Mukherjee,
  and Daniel~Benevides da~Costa,
\newblock ``A general wideband non-stationary stochastic channel model for
  intelligent reflecting surface-assisted {MIMO} communications,''
\newblock {\em IEEE Trans. Wireless Commun.}, vol. 20, no. 8, pp. 5314--5328,
  2021.

\bibitem{zhang2021joint}
Zijian Zhang and Linglong Dai,
\newblock ``A joint precoding framework for wideband reconfigurable intelligent
  surface-aided cell-free network,''
\newblock {\em IEEE Trans. Signal Process.}, vol. 69, pp. 4085--4101, 2021.

\bibitem{zheng2019intelligent}
Beixiong Zheng and Rui Zhang,
\newblock ``Intelligent reflecting surface-enhanced {OFDM}: Channel estimation
  and reflection optimization,''
\newblock {\em IEEE Wireless Commun. Lett.}, vol. 9, no. 4, pp. 518--522, 2019.

\bibitem{koziel2013surrogate}
Slawomir Koziel and Leifur Leifsson,
\newblock {\em Surrogate-based modeling and optimization},
\newblock Springer, 2013.

\bibitem{nerini2024localized}
Matteo Nerini, Golsa Ghiaasi, and Bruno Clerckx,
\newblock ``Localized and distributed beyond diagonal reconfigurable
  intelligent surfaces with lossy interconnections: Modeling and
  optimization,''
\newblock {\em arXiv:2402.05881}, 2024.

\end{thebibliography}

\end{document}